\documentclass[12pt, prd, showpacs,showkeys,superscriptaddress,nofootinbib]{revtex4}
\usepackage{amssymb}
\usepackage{amsmath}
\usepackage{setspace}

\usepackage{color}
\definecolor{DarkGreen}{rgb}{0,0.4,0}
\definecolor{DeepBlue}{rgb}{0,0,.4}
\definecolor{deepred}{rgb}{.5,0,0}
\usepackage[naturalnames=true,colorlinks=true,citecolor=DarkGreen,urlcolor=blue,linkcolor=DeepBlue]{hyperref}

\begin{document}

\title{Ba\~{n}ados-Silk-West effect with nongeodesic particles:\\
extremal horizons}
\author{I. V. Tanatarov}
\email{igor.tanatarov@gmail.com}
\affiliation{Kharkov Institute of Physics and Technology,\\
1 Akademicheskaya, Kharkov 61108, Ukraine}
\affiliation{Department of Physics and Technology, Kharkov V.N. Karazin National
University, 4 Svoboda Square, Kharkov 61022, Ukraine}
\author{O. B. Zaslavskii}
\email{zaslav@ukr.net}
\affiliation{Department of Physics and Technology, Kharkov V.N. Karazin National
University, 4 Svoboda Square, Kharkov 61022, Ukraine}

\begin{abstract}
The Ba\~{n}ados-Silk-West (BSW) effect consists in the possibility to obtain arbitrarily large energy $E_{c.m.}$ in the centre of mass frame of two colliding particles near the black hole horizon. One of the common beliefs was that the action of force on these particles (say, due to gravitational radiation) should necessarily restrict the growth of $E_{c.m.}$. We consider extremal horizons and develop a model-independent approach and analyze the conditions for the force to preserve or kill the effect, using the frames attached both to observers orbiting the black hole and to ones crossing the horizon. We argue that the aforementioned expectations are not confirmed. Under rather general assumptions, the BSW effect survives. For equatorial motion it is only required that in the proper frame the radial component of the force be finite, while the azimuthal one tend to zero not too slowly. If the latter condition is violated, we evaluate $E_{c.m.}$, which becomes indeed restricted but remains very large for small forces. 
\end{abstract}

\maketitle

\newpage \begin{spacing}{1.2}
	\tableofcontents
\end{spacing}

\newpage

\section{Introduction}

Recently, an interesting effect was discovered by Ba\~{n}ados, Silk
and West \cite{ban}, called usually the BSW effect after the names of the
authors: if two particles collide near the black hole horizon, the energy $%
E_{c.m.}$ in their centre of mass frame can grow indefinitely large,
provided the parameters of one of the particles are fine-tuned. Immediately
after this observation, several considerations of theoretical nature were
brought forward suggesting that there must be restrictions that would
prevent the realization of this effect. One of the basic objections is
connected with the force of gravitational radiation acting on particles. It
was pushed forward in \cite{berti} and is mentioned from time to time in
consequent works starting from \cite{ted}. There are also other similar effects which seem to
restrict the divergence of $E_{c.m.}$ -- say, synchrotron radiation by
charged particles near black holes \cite{fr}.

Meanwhile, the influence of the force of gravitational radiation (or any
other force) on the BSW effect is not so obvious. First of all, the BSW
effect is prepared from two main ingredients -- the presence of the horizon
and the presence of special "critical" trajectories, (see below). It was
shown in \cite{gc}, with minimal assumptions, that even for neutral
particles and nongeodesic motion, such trajectories do exist. Therefore, the
question is whether or not the force destroys these trajectories. If this
happens, the BSW effect is restricted. However, for a weak force, one can expect a large bound on $E_{c.m.}$. For instance, the analysis of
particle's motion on the innermost stable orbit near the Kerr black hole
with gravitational radiation taken into account showed that $E_{c.m.}$ can
be far beyond the Planck energy for collision of dark matter particles near
a stellar mass near-extremal black hole \cite{insp}. The analysis suggested in \cite{insp}, however, concerns special (although important for astrophysics) cases: it applies to
near-extremal Kerr black holes when fine-tuning required for the BSW effect
is realized on circular orbits. It also remains incomplete since not all
factors responsible for the self-force are taken into account. Meanwhile, it
is of interest to elucidate the issue under discussion in a
model-independent way.

In this paper we develop such a general approach and analyze the BSW effect under the influence of a generic force near the horizon of a generic axially symmetric stationary ``dirty'' black hole (i.e. a black hole that is surrounded by matter, so its metric may deviate from the Kerr one). Here we only consider the case of an extremal horizon of a maximally rotating black hole. The approach used is applicable, with minimal modifications, to static or charged black holes, as shown explicitly for the case of Reissner-Nordstr\"{o}m metric.

We consider the conditions the force should satisfy for the effect to be either preserved in some form or not. The analysis is made in terms of tetrad components of the corresponding quantities in the frames attached both to an observer orbiting the black hole, and the one crossing the horizon. The nature of the force itself is not specified, we only assume that its tetrad components in the particle's proper frame are finite and restrict our consideration to equatorial motion. 
We show that the BSW effect survives any force that satisfies the following assumptions: (i) it remains finite near the horizon, and (ii) its azimuthal component tends to zero fast enough (more detailed definition is given below). In case the above condition is not satisfied, e.g. the azimuthal force does not vanish in the horizon limit, the weaker version of the effect is realized whenever the acceleration's amplitude is small enough (as should be for e.g. radiation reaction). For the latter case, we find generic bounds on $E_{c.m.}$.

It is worth stressing that the BSW effect reveals itself not only for extremal black holes, but also for nonextremal ones. The mechanism in the latter case, however, is generally different, as it requires multiple scattering, which for extremal black holes is not necessary  \cite{gp} (see also \cite{prd}). Correspondingly, we postpone consideration of the BSW effect with a force near nonextremal horizons and, in the present paper, restrict ourselves to the extremal case. The effect for near-extremal horizons, considered in \cite{insp}, occupies an intermediate position between the two. This problem contains some subtleties on its own related to the properties of near-circular orbits and in the general setting also needs separate treatment.

There are two aspects of the BSW effect --- the behavior of $E_{c.m.}$ near the horizon and the properties of energies of the collision outcome measured at infinity. The typical energies at infinity are quite modest even in the absence of force \cite{inf1,inf2,inf3}, so taking the force into account can only change them slightly. It is the first aspect which is nontrivial and is being discussed in the present paper. 

The paper is organized as follows. In Sec. II, we consider classification of
particles relevant for the BSW effect and discuss novel features that the
force brings into the system. In Sec. III, we consider behavior of
acceleration near the horizon in different frames (attached to an observer orbiting the black hole or to one crossing the horizon). In Sec. IV, we illustrate
general relationships using the Reissner-Nordstr\"{o}m metric as an example.
In Sec. V, we consider generic motion in the equatorial plane under the action of finite forces and derive conditions on the force that allow or forbid critical trajectories.
In Sec. VI, we estimate the bounds on $E_{c.m.}$ for the case when the force is least favourable for the effect but small. In Sec. VII, we discuss pure kinematic restrictions on particle's trajectories (valid even in the absence of force) which can influence the properties of the BSW effect. Sec. VIII is devoted to conclusion.

\section{Particles' kinematics near extremal horizons}

\subsection{A particle in axially symmetric metric}

We consider the axially symmetric stationary metric written (at least in the
vicinity of the horizon) in coordinates which are obtained from the Gaussian
normal ones by replacing the distance to the horizon $n$ with the
radial coordinate $r$, defined so that\footnote{Such as the quasiglobal coordinate of \cite{BrRub}, ch.3.} $A(r)\sim N^{2}$ in the
horizon limit, where $N^{2}\rightarrow 0$ (hereafter $c=1$): 
\begin{equation}
ds^{2}=-N^{2}dt^{2}+g_{\phi }(d\phi -\omega dt)^{2}+\frac{dr^{2}}{A}%
+g_{z}dz^{2}.
\end{equation}

Let there be some arbitrary, not necessarily geodesic, particle of mass $m$,
four-velocity $u^{\mu }$ and four-momentum 
\begin{equation*}
p^{\mu }=mu^{\mu }.
\end{equation*}%
It is convenient to represent the four-velocity, both with upper and lower
indices, by the components of its four-momentum in the following way: 
\begin{align}
u^{\mu }& =\frac{1}{m}\Big(\frac{X}{N^{2}},\frac{L}{g_{\phi }}+\frac{\omega X%
}{N^{2}},p^{r},p^{z}\Big);  \label{Uu} \\
u_{\mu }& =\frac{1}{m}\Big(-E,L,\frac{1}{A}p^{r},g_{z}p^{z}\Big),  \label{Ud}
\end{align}%
where $E=-mu_{0}$ is energy, $L=mu_{\phi }$ angular momentum and 
\begin{equation}
X=E-\omega L.  \label{x}
\end{equation}
Due to forward in time condition, $X$ is always positive.

For a free particle on a geodesic trajectory the energy $E$ and angular momentum 
$L$ are conserved; eqs. (\ref{Uu}), (\ref{Ud}) are nothing but the equations
of motion with given fixed values of $E$ and $L$. In the general case, $E$
and $L$ are not conserved and together with $u^{z}$ should be treated as
functions of the particle's proper time. Nonetheless, we still write the
components of the four-velocity in the same form (\ref{Uu}), (\ref{Ud})
which can be considered simply as useful parametrization.

The normalization condition $u^\mu u_\mu =-1$ can be written as 
\begin{equation}
\frac{1}{A}(p^r)^2 +g_{z}(p^z)^2 =\frac{X^2}{N^2}-\frac{L^2}{g_\phi}-m^2 .  \label{norm}
\end{equation}

Then $p^{r}$ is expressed through the three independent parameters $E$, $L$
and $u^{z}$: 
\begin{equation}
p^{r}=\pm \frac{\sqrt{A}}{N}\;Z,  \label{pr}
\end{equation}
where 
\begin{align}
Z^{2} 
& =X^{2}-N^{2}\Big[\frac{L^{2}}{g_{\phi }}+g_{z}(p^{z})^{2}+m^{2}\Big]. \label{z}
\end{align}

The formulas in this section are applicable also to massless particles, with
the only difference that one has to set $m=0$. The four-momentum then, in
the appropriate parametrization of the worldline, is related to the wave
vector $k^{\mu }$ as $p^{\mu }=\hbar k^{\mu }$.

\subsection{Two particles' collision near horizon}

The energy $E_{i\;c.m.}$ of a particle $i$ with four-momentum $p_{i}^{\mu }$
in its center of mass (c.m.) frame is simply its rest mass, i.e. the norm of
its four-momentum: 
\begin{equation}
E_{i\;c.m.}^{2}=m_{i}^{2}=-p_i^{\mu }p_{i\,\mu }.
\end{equation}%
Likewise, for two particles with masses $m_{1}$ and $m_{2}$ and
four-velocities $u_{1}^{\mu }$ and $u_{2}^{\mu }$ the center of mass energy $%
E_{c.m.}$ at the collision event is the norm of their total four-momentum 
\begin{equation}
E_{c.m.}^{2}=-(p_{1}^{\mu }+p_{2}^{\mu })(p_{1\mu }+p_{2\mu
})=m_{1}^{2}+m_{2}^{2}+2m_{1}m_{2}\gamma _{c.m.},  \label{cm}
\end{equation}%
where 
\begin{equation}
\gamma _{c.m.}=-u_{1\mu }u_{2}^{\mu }
\end{equation}%
is the relative Lorentz factor.

The contraction can be written as 
\begin{equation}
m_{1}m_{2}\gamma _{c.m.}=\frac{X_{1}X_{2}-Z_{1}Z_{2}}{N^{2}}-\frac{L_{1}L_{2}%
}{g_{\phi }}-g_{z}p_{1}^{z}p_{2}^{z}.  \label{ga}
\end{equation}

For a collision of a massive particle of mass $m$ and a photon one obtains
that 
\begin{equation}
E_{c.m.}^{2}=m^{2}+2m\hbar \omega _{det},
\end{equation}%
where $\omega _{det}$ $=-k_{\mu }u^{\mu }$ is the photon's frequency as
detected in the frame of this massive particle.

\subsection{Usual and critical particles near extremal horizons}

Consider a particle in the vicinity of a regular extremal horizon, for which 
\cite{reg}
\begin{align}
N^{2}(r)&\sim (r-r_{H})^{2},  \label{Extr}\\
\omega (r)&=\omega _{H}-\omega _{1}(z)N+O(N^{2}),\qquad \omega _{H}=const,
	\label{OmExpansion}
\end{align}
where $r=r_{H}$ is the horizon. The regularity of the horizon implies \cite{reg} that,
in particular, $\omega_H$ is a constant, and that other metric functions can
also be expanded into series by $(r-r_{H})$ with positive powers. Note also
the sign by $\omega_1$, defined so for consistency with earlier works.

From the normalization condition (\ref{norm}) then a particle's
four-velocity components can always be presented as series by $N$ (though
they can diverge at the horizon). Assuming $L$ and $E$ are finite, 
\begin{equation}
E=E_{H}+E_{1}N+O(N^{2}),\quad L=L_{H}+L_{1}N+O(N^{2}),  \label{el}
\end{equation}%
and then 
\begin{align}
X=& X_{H}+O(N), \\
& X_{H}=E_{H}-\omega _{H}L_{H}.
\end{align}

For a usual (generic) particle $X_{H}\neq 0$. The normalization (\ref{norm})
then implies that 
\begin{equation}
u^{r}=O(1),  \label{ru}
\end{equation}%
$u^{z}=O(1/N)$, so the particle reaches the horizon in finite proper time $%
\tau \sim \int dr<\infty $.

However, there are also worldlines of particles with angular momentum
fine-tuned to energy in such a way that $X_{H}=0$, so that 
\begin{equation}
X=O(N).  \label{xc}
\end{equation}%
Such particles are called critical.

For critical particles the right hand side of normalization condition(\ref{norm}) is bounded, and as the left hand side there is a sum of squares, we obtain 
\begin{equation}
u^r =O(N),\quad u^z =O(1).  \label{Ubounded}
\end{equation}

Then the equation of radial motion in the main order by $N$ is 
\begin{equation}
\frac{dr}{d\tau }=-\frac{r-r_{H}}{\tau _{0}},  \label{rt}
\end{equation}%
where $\tau _{0}$ is a constant for motion in equatorial plane\footnote{In general, the coordinate $z$ can oscillate between some limiting values, see \cite{ne} for the Kerr metric and  \cite{jh} for discussion of a more general case.}; its solution is 
\begin{equation}
r-r_{H}=r_{0}e^{-\tau /\tau _{0}}  \label{ur}
\end{equation}%
and the proper time of reaching the horizon diverges as $\ln (r-r_H )$. In case $u^{r}$ is of higher order that $N$, the divergence is stronger (i.e. if $u^{r}\sim (r-r_{H})^2$,
then $\tau $ diverges as $(r-r_{H})^{-1}$).

\subsubsection{Example: the Kerr metric}

It is instructive to look at the critical trajectory for the Kerr metric.
Let us restrict ourselves to equatorial motion $\theta =\frac{\pi }{2}$.
Then, the metric coefficients near the horizon of the extremal Kerr black
hole read 
\begin{align}
&N\approx \frac{r-r_{H}}{2r_{H}},\qquad \sqrt{A}\approx \frac{r-r_{H}}{r_{H}}%
,\qquad (g_{\phi })_{H}=4r_{H}^{2}, \\
&\omega _{H}=\frac{1}{2r_{H}},\qquad \omega -\omega _{H}\approx -\frac{%
r-r_{H}}{2r_H^2}.
\end{align}
Then, it follows from eqs. (\ref{pr}), (\ref{z}) that the trajectory of the
particle with $E=\omega _{H}L$ has exactly the form (\ref{rt}) with 
\begin{equation}
\tau _{0}=\frac{r_{H}}{\sqrt{3\frac{E^{2}}{m^{2}}-1}}.
\end{equation}

\subsection{The BSW effect}

Consider the collision of two particles. For a usual (generic) particle,
assuming $E$, $L$ and $u^z$ are finite\footnote{Those are natural assumptions, but for justification see the section on
dynamics below.}, 
\begin{equation}
X=X_H +O(N),\quad Z=X+O(N^2).
\end{equation}

Then the relative Lorentz factor at the collision event of two usual
particles is 
\begin{equation}
m_{1}m_{2}\gamma _{c.m.}=\frac{X_{1}X_{2}-Z_{1}Z_{2}}{N^{2}}+O(1)=O(1).
\label{gacm}
\end{equation}

However, for a critical particle 
\begin{equation}
X=X_{N}N+O(N^{2}),\quad Z=Z_{N}N+O(N^{2}).
\end{equation}%
Then for two critical particles $\gamma _{c.m.}$ is also bounded, but the
relative Lorentz factor at the collision event of a critical (1) particle
and a usual (2) particle is 
\begin{equation}
m_{1}m_{2}\gamma _{c.m.}=\frac{X_{H}^{(2)}(X_{N}^{(1)}-Z_{N}^{(1)})}{N}%
+O(1)\rightarrow \infty .  \label{div}
\end{equation}

So, the BSW effects occurs whenever one usual and one critical particle collide near the horizon. Geodesic particles can be
critical just due to the choice of initial conditions which fix $E$ and $L$,
so one can always achieve $X_{H}=0.$ The question is how resilient is the
criticality attribute with respect to acceleration: whether a particle can
remain critical under the action of finite forces, such as radiation
reaction.

\subsection{Generalization: usual, critical and sub-critical particles}

In the absence of external forces acting on a particle, in the vicinity of a
regular horizon, where all metric functions can be expanded into series by
the radial coordinate $r$, the geodesic equation induces the same
type of expansions for the parameters of a particle, such as $X$ and $E$.
Therefore there are only two principally different types of particles: usual
and critical ones. If we want to take into account forces acting on a
particle, however, we have to allow for more general setting. In particular,
we assume that acceleration components in the proper frame of a particle and 
$X$ can behave as $\xi^{q}$ and $\xi^{p}$ respectively, with some real $q$ and $p$, where
\begin{equation}
 \xi \equiv r-r_{H}. \label{xi}
\end{equation}
Hereafter we consider this reasonably general while still relatively simple model.

If a particle reaches the horizon, $X$ must tend to zero more slowly than $N$, so that $Z^2$ remains positive. This is only possible for $p\leq 1$. On the other hand, it is reasonable to
restrict our consideration to finite $E$ and $L$, and thus $X$, so $p\geq 0$. Then there are three possible particle types, distinguished by $p$ in 
\begin{equation}
X\sim \xi^ p .
\end{equation}

\begin{enumerate}
\item $p=0$: \textbf{usual particles}. 
\begin{equation}
X_u=x_H + x_1 \xi +\ldots, \qquad Z_u=X+O(\xi^2).
\end{equation}

\item $p=1$: \textbf{critical particles}. 
\begin{equation}
X_{cr}=x_1 \xi + x_2 \xi^2 +\ldots,\qquad Z_{cr}=O(\xi) .
\end{equation}

\item $p\in (0,1)$: the intermediate case, which will be called \textbf{sub-critical} particles hereafter: 
\begin{equation}
X_{sc}=\alpha \xi^{p}(1+x_1 \xi+\ldots),\qquad Z_{sc}=X_{sc}+O(\xi^{2-p}).
\end{equation}
Their proper time of reaching the horizon $\sim \int d\xi /Z_{sc}$ is finite.
\end{enumerate}

As shown above (\ref{ga}), for collision of two particles the relative Lorentz factor
is 
\begin{equation}
\gamma_{c.m.} =\frac{X_1 X_2 -Z_1 Z_2}{m_1 m_2 N^2}+O(1).  \label{RelGamma}
\end{equation}
For collision of two usual or two critical particles near extremal horizon, for which $N^2 \sim
\xi^2$ (\ref{Extr}), we have $\gamma_{c.m.} =O(1)$; for usual and critical $\gamma_{c.m.} \sim 1/\xi$.
Likewise for usual and sub-critical one obtains 
\begin{equation}
\gamma_{c.m.}\sim \xi^{-p} \to \infty ;
\end{equation}
for critical and sub-critical 
\begin{equation}
\gamma_{c.m.} \sim \xi^{p-1}\to \infty .  \label{SubGamma}
\end{equation}
So, the corresponding particles behave as
critical in collisions with usual ones and as usual in collisions with
critical ones.

This result can be derived in the general setting. First of all, let there
be a particle with 
\begin{align}
&X=\alpha \xi^p (1+O(\xi)),\qquad p\in [0,1]; \label{alpha} \\
& \frac{L^2}{g_{\phi}}+m^2 =\beta^2+O(\xi),\qquad \alpha,\beta\sim 1. \label{beta}
\end{align}

Then 
\begin{equation}
Z-X =-C\xi^{2-p}(1+O(\xi)),  \label{Z-X}
\end{equation}
where 
\begin{equation}
C=\left\{\begin{array}{ll}
	\beta^2 / 2\alpha & \quad \text{for}\quad p< 1; \\ 
	\alpha-\sqrt{\alpha^2 -\beta^2} & \quad \text{for}\quad p=1 .
\end{array}\right.
\end{equation}
As due to forward in time condition $\alpha>0$ and $\beta^2$ is also positive, $C$ is strictly positive as well.

Now suppose we have two such particles, with $p_{1}$ and $p_{2}$, colliding
near the horizon. Then using (\ref{RelGamma}) and (\ref{Z-X}), the relative
Lorentz factor is reduced to 
\begin{align}
m_{1}m_{2}\,\gamma _{c.m.}& 
	=O(1)+\big(1+O(\xi )\big)\Big[C_{1}\alpha _{2}\xi^{p_{2}-p_{1}}
	+C_{2}\alpha _{1}\xi ^{p_{1}-p_{2}}\Big]  \label{gamma12-1} \\
& \sim \xi ^{-|p_{1}-p_{2}|}\big[1+O(\xi )+O(\xi ^{2|p_{1}-p_{2}|})\big],
\label{gamma12-2}
\end{align}%
and thus 
\begin{equation}
\gamma_{c.m.}\sim \xi ^{-|p_{1}-p_{2}|}.  \label{p1-p2}
\end{equation}%
Here, gamma becomes finite only if $p_{1}=p_{2}$. We see that consideration
of sub-critical particles is convenient, as it allows to describe usual and
critical particles in a more coherent and unified way, while at the same
time providing greater generality, necessary when dealing with non-geodesic
motions.

\section{Dynamics}

\subsection{OZAMO and FZAMO frames}

There are two main qualitatively different frames of reference in the
vicinity of a black hole horizon. The tetrad vectors and tetrad components
of different quantities will be denoted by superscripts in parenthesis,
while low case ``o'' or ``f'' in the subscript will denote which frame is used, i.e. $%
a_{o}^{(t)}$ is the $t$-component of acceleration in the OZAMO frame (see
below).

\paragraph{OZAMO.}

The first kind of frame is attached to an observer who is orbiting the black hole with
constant $r$, having constant energy and zero angular momentum. We will call
it OZAMO for orbital zero angular momentum observer\footnote{This observer is usually called just ZAMO in textbooks, but we need to be more specific.}. It is the analogue of
the static observer in a static spacetime, and it becomes lightlike in the
horizon limit \cite{72}.

The tetrad 1-forms of the OZAMO frame, denoted by small ``o'' subscripts,
read 
\begin{align}
e^{(t)}_o& =-Ndt;  \label{zamo} \\
e^{(\phi )}_o& =g_{\phi }^{1/2}\;(d\phi-\omega dt); \\
e^{(r)}_o& =A^{-1/2}dr; \\
e^{(z)}_o& =g_{z}^{1/2}dz.  \label{zamo-z}
\end{align}

If another particle's four-velocity is $u^{\mu }$, then its Lorentz factor
in this frame is 
\begin{equation}
\gamma=-u^{\mu }(e_{o }^{(t)})_\mu=\frac{X}{mN}.  \label{gx}
\end{equation}
Thus for a particle with $X\sim \xi^p$
\begin{equation}
\gamma \sim \xi^{p-1}; \label{gx1}
\end{equation}
for a usual particle it diverges in the horizon limit, while for a critical
one it stays finite.

A particle's acceleration is 
\begin{equation}
a^{\mu }\equiv u^{\nu }\nabla _{\nu }u^{\mu }.
\end{equation}
Its tetrad components in the OZAMO frame 
\begin{equation*}
a^{(i)}_o=a^\mu (e^{(i)}_o)_{\mu},\qquad i=t,\phi,r,z,
\end{equation*}
are equal to 
\begin{align}
a^{(t)}_o& =Na^{t};  \label{a} \\
a^{(\phi) }_o& =\sqrt{g_{\phi }}\;(a^{\phi }-\omega a^{t}); \label{a-phi} \\
a^{(r)}_o& =\frac{1}{\sqrt{A}}\;a^{r}; \label{a-r}\\
a^{(z)}_o& =\sqrt{g_{z}}a^{z}; \label{a-z}
\end{align}%
the acceleration scalar then can be presented as 
\begin{equation}
a^{2}\equiv a^{\mu }a_{\mu } =-(a^{(t)}_o)^{2} +(a^{(\phi)}_o)^{2}
+(a^{(r)}_o)^{2}+(a^{(z)}_o)^{2}.
\end{equation}

The OZAMO orbits the horizon at constant $r$ and does not cross it,
therefore it is not classified as either usual or critical particle, which
does cross or approach the horizon in infinite proper time respectively. However, it
is useful to note, that, as its Lorentz factor is finite with respect to a
critical particle, and vice versa, in the discussion that follows, OZAMO and
critical particles behave similarly.

It is well-known, that an OZAMO frame breaks down at the horizon, thus strictly speaking at the horizon it is not a valid frame and OZAMO is not an observer in the traditional sense. Hereafter, what we refer to as the values of some quantities measured in the OZAMO frame in the horizon limit are the limits of the corresponding quantities measured in successive different OZAMO frames, with different $r_{ZAMO}$, when $r_{ZAMO}\to r_H$.

\paragraph{FZAMO.}

The other important frame of reference is realized by one of the usual
particles crossing the horizon. For simplicity, it is convenient to take for
such an observer $L=0$ similarly to OZAMO and, additionally, $E=m$. Thus we
will call the corresponding observer FZAMO for falling zero angular
momentum observer. Its frame $\{e_{f}^{(i)}\}$, with $i=t,\phi ,r,z$, is
constructed by making a local Lorentz transformation from the OZAMO in the
direction towards the horizon\footnote{Note that this is the transformation for one-forms; vectors are
transformed by the inverse matrix, which differs by the sign of $v_{f}$.}: 
\begin{align}
\begin{pmatrix}
(e^{(t)}_{f})_{\mu} \\ 
(e^{(r)}_{f})_{\mu}%
\end{pmatrix}
&=\gamma_f
\begin{pmatrix}
1 & v_f \\ 
v_f & 1%
\end{pmatrix}
\begin{pmatrix}
(e^{(t)}_o)_{\mu} \\ 
(e^{(r)}_o)_{\mu}
\end{pmatrix}
\label{rrt}
\end{align}%
The FZAMO's Lorentz factor in the OZAMO frame is $\gamma _{f}=-\left( u^{\mu
}\right) (e_{o}^{(t)})_{\mu }$, where $u^{\mu }$ is given by (\ref{Uu}) with $L=0$ and $E=m$, and $v_{f}=\sqrt{1-\gamma_{f}^{-2}}$.

Then, 
\begin{equation}
\gamma _{f}=\frac{1}{N},\qquad v_{f}=\sqrt{1-N^{2}}.  \label{va}
\end{equation}

The corresponding tetrad components of acceleration 
\begin{equation}
a_{f}^{(i)}=a^{\mu }(e_{f}^{(i)})_{\mu },\qquad i=t,\phi ,r,z  \label{aa}
\end{equation}%
are related to $a_{o}^{(i)}$ by the respective Lorentz transformation
which becomes singular on the horizon, where $N\rightarrow 0$, $\gamma
_{f}\rightarrow \infty $.

\paragraph{Proper frame.}

For non-critical particles with $p<1$ the Lorentz factor relative to the
OZAMO frame diverges as $\gamma \sim \xi ^{p-1}$ (\ref{SubGamma}). Thus the
correct reference frame for it will have the same behaviour of Lorentz
factor. We will construct it, analogously to FZAMO, by making the
corresponding boost in the radial direction, and call it for simplicity the
proper frame for a particle, although it may not be exactly proper. What is
important is that, in contrast to the OZAMO, the particle's velocity in it
stays finite (does not tend to $c$).

Thus, given a particle's Lorentz factor in the OZAMO frame, $\gamma$
(without subscripts), the tetrad components of acceleration in the proper
frame are 
\begin{equation}
\begin{pmatrix}
a^{(t)}_{pr} \\ 
a^{(r)}_{pr}%
\end{pmatrix}
=\gamma 
\begin{pmatrix}
1 & v \\ 
v & 1%
\end{pmatrix}
\begin{pmatrix}
a^{(t)}_o \\ 
a^{(r)}_o%
\end{pmatrix}%
.  \label{AProper}
\end{equation}
For a usual or critical particle this reduces to the already considered
OZAMO and FZAMO frames respectively, while for sub-critical particles the proper frame
does not coincide with either one of those.

\subsection{Acceleration in different frames}

When describing particles' motion near the horizon, we must restrict
ourselves to particles with finite acceleration. This necessarily means that
the acceleration scalar $a^{2}$ should be finite. It would seem that it is
natural to demand that tetrad components of acceleration are finite as well.
However, as shown above, when we describe a particles' motion near the
horizon, we have different frames of reference, which are related to each
other by singular Lorentz transformations. This means that finite tetrad
components of acceleration in one of the frames may correspond to diverging
tetrad components in the other or vice versa.

The frame in which tetrad components of a particle's acceleration should be
finite is the instantly comoving frame, or equivalently, any frame which
moves with finite Lorentz factor with respect to that. For example, recall
the reasonably realistic problem of a charged particle in a uniform electric
field in Special Relativity. The tetrad components of acceleration in the
laboratory frame (with Minkowski metric) diverge proportionally to the
Lorentz factor, while those in the instantly comoving frame (with the
Rindler metric and the horizon) are constant (see, e.g. p. 403 of \cite{car}%
).

For a critical particle then acceleration is adequately and most easily
measured in the OZAMO frame. For a usual particle we would have to attach
the tetrad also to one of the usual particles, for example to FZAMO. Due to
normalization of four-velocity $u^\mu u_\mu =-1$, which implies $a^\mu u_\mu
=0$, in each case it is sufficient to show that three of the four
tetrad components are finite.

\subsection{Energy and angular momentum}

If $\xi ^{\mu }$ is a Killing vector field, then 
\begin{equation}
\frac{d}{d\tau }(\xi ^{\mu }u_{\mu })=\xi ^{\mu }a_{\mu }.
\end{equation}%
In a stationary axisymmetric metric we have two Killing vectors $\xi
_{t}^{\mu }=\delta _{t}^{\mu }$ and $\xi _{\phi }^{\mu }=\delta _{\phi
}^{\mu }$, which give 
\begin{align}
\frac{1}{m}\frac{dE}{d\tau }& =(N^{2}-\omega ^{2}g_{\phi })a^{t}+\omega
g_{\phi }a^{\phi }; \\
\frac{1}{m}\frac{dL}{d\tau }& =-\omega g_{\phi }a^{t}+g_{\phi }a^{\phi },
\end{align}%
or through the tetrad components in the OZAMO frame (\ref{a}, \ref{a-phi})
\begin{align}
\frac{1}{m}\frac{dE}{d\tau }& =Na^{(t)}_o +\omega \sqrt{g_{\phi }}%
\;a^{(\phi)}_o;  \label{dEdt} \\
\frac{1}{m}\frac{dL}{d\tau }& =\sqrt{g_{\phi }}\;a^{(\phi)}_o.  \label{dLdt}
\end{align}%

It is clear, that if the proper time of crossing the horizon is finite, as
is the case for the usual particles, then the finiteness of $Na^{(t)}_o$ and 
$a^{(\phi)}_o$ implies that $E$ and $L$ are also bounded. However, this does
not seem to be necessarily so for critical particles, for which the proper
time of reaching the horizon diverges.

\subsection{Dynamic restrictions on a particle's velocity}

Let us enumerate and classify all the possible variants of particle's type
of asymptotic motion in the vicinity of the horizon, now in more detail than
in the section on kinematics, so as to focus below only on those that are
not explicitly non-physical. 

First of all, diverging $L$, as seen from (\ref{dLdt}), would correspond to
continuous acceleration in the $\phi$ direction, which would cost formally
infinitely large amounts of fuel per a unit mass particle. If one has the
resources to make such experiments, he would not need the BSW effect in
order to observe (formally) infinite energy in the center of mass frame. So
this variant is of not much interest.

Secondly, one could imagine divergent $u^z$. Such a particle would have
velocity tending to $c$ and directed along the $z$ axis (or at finite angle
with respect to it) both in the OZAMO and FZAMO frames. This would mean that the
particle is ``accelerated'' (in the sense that its velocity increases) not
only in radial direction, but also along the horizon surface. This would be
very strange behaviour, and in the Kerr metric such particles are naturally
absent \cite{ne}. We will not consider this variant here.

Given these two natural assumptions, from the normalizing condition 
\begin{equation}
Z^{2}=X^{2}-N^{2}\tilde{\beta}^{2},
\end{equation}%
where $\tilde{\beta}$ in the horizon limit tends to a positive real number,
finite and separated from zero. Consequently, for a particle reaching the
horizon, where $Z^{2}$ must remain positive, and at the same time finite $X$ and $E$ (see discussion after Eq. (\ref{xi})), we have
\begin{equation}
X\sim \xi ^{p},\quad \text{with}\quad p\in[0,1],
\end{equation}%
which corresponds to usual, sub-critical and critical particles as discussed
above.

\subsection{Usual particles}

For a usual particle $X_{H}\neq 0$ by definition. As discussed above, the
tetrad components of its acceleration in the FZAMO frame $a^{(i)}_f$ must be
finite. Then the components in the OZAMO frame $a^{(i)}_o$, related to them
via the singular Lorentz transform (\ref{va}), with $\gamma_f
=1/N$, can diverge as $1/N$. Writing out explicitly the asymptotics for the $%
t$ and $r$ components in both frames, we get 
\begin{align}
a^{(t)}_f &= (a^{(t)}_f)_{0}+(a^{(t)}_f)_1 N+O(N^2);  \label{af-t} \\
a^{(r)}_f &= (a^{(r)}_f)_{0}+(a^{(r)}_f)_1 N+O(N^2);  \label{af-r} \\
a^{(t)}_o &=+\frac{ (a^{(t)}_f)_{0}- (a^{(r)}_f)_{0}}{N} +\big[ %
(a^{(t)}_f)_{1}-(a^{(r)}_f)_{1}\big]+O(N);  \label{ao-t} \\
a^{(r)}_o &=-\frac{ (a^{(t)}_f)_{0}- (a^{(r)}_f)_{0}}{N} -\big[ %
(a^{(t)}_f)_{1}-(a^{(r)}_f)_{1}\big]+O(N),  \label{ao-r}
\end{align}

The $\phi $ and $z$ components are the same in the two frames and must be
bounded: 
\begin{align}
a^{(\phi)}_f &=a^{(\phi)}_o =O(1); \\
a^{(z)}_f &=a^{(z)}_o =O(1).
\end{align}
Then, we see that if $a^{(i)}_f=O(1)$, the right hand side of (\ref{dEdt}) is
finite. The left hand side is also finite, as for a usual particle, given $%
u^{r}\sim 1$ (\ref{ru}), $dr\sim d\tau \sim dN$.

The explicit expressions for $a_{o}^{(r)}$ and $a_o^{(z)}$ are 
\begin{align}
a_o^{(r)}=& \frac{1}{\sqrt{A}}\Big\{(u^{r}\partial _{r}+u^{z}\partial
_{z})u^{r} -\frac{A^{\prime }}{2A}(u^{r})^{2} -\frac{A}{2}\partial
_{r}g_{z}(u^{z})^{2}+  \notag \\
& -\frac{A}{2}\Big[X^2 \partial_r N^{-2}-L^2 \partial_r g_\phi^{-1}
	-2\frac{XL}{N^2}\partial_r \omega\Big]\Big\};  \label{aR}
\\
a_o^{z}=& \sqrt{g_z}\Big\{ (u^{r}\partial _{r}+u^{z}\partial _{z})u^{z} +%
\frac{\partial _{z}g_{z}}{2g_{z}}(u^{z})^{2} 
	+\frac{\partial _{r}g_{z}}{g_{z}}u^{r}u^{z}+  \notag \\
& -\frac{1}{2 g_z}\Big[X^2 \partial_z N^{-2}-L^2 \partial_z g_\phi^{-1}
	-2\frac{XL}{N^2}\partial_z \omega\Big]\Big\}.
\label{aZ}
\end{align}

The conditions $a^{(i)}_f=O(1)$ can be reformulated in the form of
restrictions on the coefficients $\alpha_k$ and $\beta_k$ in the expansions 
\begin{align}
u^{r}&=\alpha _{0}(z)+\alpha _{1}(z)(r-r_{H})+O((r-r_{H})^{2}),  \label{u} \\
u^{z}&=\beta _{0}(z)+\beta _{1}(z)(r-r_{H})+O((r-r_{H})^{2}).
\end{align}

\subsection{Critical particles}

Such particles approach the horizon but, in contrast to usual ones, the
process takes infinite proper time. On the other hand, as seen from (\ref{xc}) and (\ref{gx}), their Lorentz factor in the OZAMO frame $\gamma $ is
finite, and the velocity is $v<1$, so the tetrad components of acceleration
in the OZAMO frame must be finite. As mentioned above, we consider only
motion with $E$ and $L$ bounded in the horizon limit $\tau \rightarrow
\infty $. This means that $a_{o}^{(\phi )}$ should be not only bounded, but
integrable (\ref{dLdt}): $\int d\tau a_{o}^{(\phi )}<\infty $. If we assume
that $a_{o}^{(\phi )}$ is expandable in power series by $r$ with integer powers, this means 
\begin{equation}
a_{o}^{(\phi )}=O(N).
\end{equation}%
With this condition satisfied, and Eq. (\ref{ao-t}) taken into account, the boundedness of $E$ from (\ref{dEdt}) does
not give any more restrictions on $a_{o}^{(t)}$. Then, using that $E$, $L$,
and $u^{z}$ are bounded (\ref{el}), while $u^{r}$ and $X$ are $O(N)$, and $\omega
_{H}=const$ (which follows from regularity \cite{reg}), it is easy to see
that all the terms in (\ref{aR}) and (\ref{aZ}) are automatically finite, so 
\begin{equation}
a_{o}^{(r)},a_{o}^{(z)}=O(1).
\end{equation}
Thus all components of acceleration of a critical particle in the OZAMO
frame, and therefore in the instantly comoving proper frame, are finite
unconditionally. This is in contrast to usual particles, for which the
conditions $a_{f}^{(i)}=O(1)$ impose some additional constraints on $\alpha
_{k}$ and $\beta _{k}$ in (\ref{u}).

In the FZAMO frame, and the frame of any usual particle, the picture looks
different, as the relative Lorentz factor of a usual and critical particle
diverges as $1/N$. Using the Lorentz transformation (\ref{rrt})
between the OZAMO and FZAMO frames, with $\gamma_f \sim 1/N$, we see that $%
a^{(t)}_f$ and $a^{(r)}_f$ can diverge as $1/N$. Using (\ref{rrt}), the asymptotics of these components of acceleration in the two
frames can be brought to the form 
\begin{align}
a^{(t)}_o &= (a^{(t)}_o)_{0}+(a^{(t)}_o)_1 N+O(N^2); \\
a^{(r)}_o &= (a^{(r)}_o)_{0}+(a^{(r)}_o)_1 N+O(N^2); \\
a^{(t)}_f &=\frac{ (a^{(t)}_o)_{0}+ (a^{(r)}_o)_{0}}{N} +\big[ %
(a^{(t)}_o)_{1}+(a^{(r)}_o)_{1}\big]+O(N); \\
a^{(r)}_f &=\frac{ (a^{(t)}_o)_{0}+ (a^{(r)}_o)_{0}}{N} +\big[ %
(a^{(t)}_o)_{1}+(a^{(r)}_o)_{1}\big]+O(N).
\end{align}
The $\phi$ and $z$ components in the two frames are the same and therefore,
as shown above, satisfy 
\begin{align}
a^{(\phi)}_o &=a^{(\phi)}_f =O(N); \\
a^{(z)}_o &=a^{(z)}_f =O(1).
\end{align}

Thus we have two mutually complimentary cases. In the OZAMO frame $r$ and $t$
components of acceleration diverge for usual particles and stay finite for
the critical ones. In the FZAMO frame, the situation is opposite: $r$ and $t$
components of acceleration are finite for usual particles and diverge for
the critical ones. The $\phi$ and $z$ components are the same in the two
frames and are finite. For critical particles, additionally $%
a_{o}^{(\phi)}=O(N)$ near the horizon for energy and angular momentum to
remain bounded.

\section{Example: the Reissner-Nordstr\"{o}m metric}

The approach and results of the present paper are also valid in the case of
the electromagnetic interaction with minimal changes: in eq. (\ref{x}) one
should make the replacement $X\rightarrow X-q\varphi $, where $\varphi $ is
the electrostatic potential, and $q$ is the particle's charge. In order to
demonstrate this, it is instructive to consider as an example the extremal
Reissner-Nordstr\"{o}m metric. In this case the metric functions are 
\begin{equation}
N=\sqrt{A}=1-\frac{r}{r_{H}},\qquad \omega =0,\qquad g_{\phi }=r^{2},
\end{equation}%
and the electromagnetic field potential is 
\begin{equation}
A_{\mu }=-\varphi \delta _{\mu }^{t},\qquad \varphi =\frac{Q}{r},
\end{equation}%
where 
\begin{equation*}
Q=r_{H}
\end{equation*}%
is the extremal black hole's charge, so that the only nonvanishing
components of the electromagnetic field tensor are 
\begin{equation*}
F_{rt}=-F_{tr}=\frac{Q}{r^{2}}.
\end{equation*}

For a particle of charge $q$ moving radially towards the horizon the
four-momentum can be parametrized as 
\begin{equation}
p_\mu=m u_\mu=-(X,0,Z/N^2 ,0) ,
\end{equation}
then the normalization condition implies 
\begin{equation}  \label{zz}
Z=\sqrt{X^2 -m^2 N^2}.
\end{equation}
The equation of motion 
\begin{equation}
ma^{\mu }=qF^{\mu \nu }u_{\nu },
\end{equation}
has the integral of motion 
\begin{equation}  \label{RN-energy}
E=X+q\varphi =const.
\end{equation}

For usual particles, with $X_H \neq 0$, 
\begin{equation}
p^{r}=-X_{H}-\frac{q}{r_H}(r-r_{H})+O((r-r_{H})^{2}),
\end{equation}%
in agreement with (\ref{u}).

For a critical particle
\begin{equation}
E=q,\quad X=qN,\quad Z=N\sqrt{q^{2}-m^{2}}.\label{RNcrit}
\end{equation}
Then, integrating the equation for radial motion 
\begin{equation*}
\frac{dr}{d\tau}=-\frac{Z}{m} =-N(r)\sqrt{q^2 /m^2 -1},
\end{equation*}
it is easy to obtain that in the horizon limit the same asymptotic as in eq.
(\ref{ur}) holds, with the characteristic time 
\begin{equation}
\tau _{0}=r_H \Big(\frac{q^2}{m^2}-1\Big)^{-1/2}.
\end{equation}

Now, we will consider the acceleration measured by the two types of
observers.

\subsection{Static observers}

The tetrad (\ref{zamo})--(\ref{zamo-z}) in this case turns into the tetrad
of a static observer. Then, using (\ref{zamo}) and (\ref{a}), we obtain 
\begin{align}
m a^{(t)}_o&=-\frac{qQ}{r^{2}}\,\frac{Z}{mN}, \\
m a^{(r)}_o&=+\frac{qQ}{r^{2}}\,\frac{X}{mN}, \\
&m^2 a^{2}=\Big(\frac{qQ}{r^2}\Big)^2 .
\end{align}

For the critical particle (\ref{RNcrit}) both components of acceleration 
\begin{align}
m a^{(t)}_o&=-\frac{qQ}{r^{2}}\sqrt{q^{2}/m^{2}-1}, \\
m a^{(r)}_o&=\frac{qQ}{r^{2}}\frac{E}{m}
\end{align}
are finite on the horizon, and can be expanded into a series by $(r-r_{H})$
or $N$.

However, for a usual particle, with $X_{H}\neq 0$, near the horizon 
\begin{equation*}
X=X_{H}+O(N),\quad Z=X+O(N^2),
\end{equation*}
so 
\begin{align}
&a^{(r)}_o \approx -a^{(t)}_o= \frac{a_{-1}}{N}+O(1),
\end{align}
where 
\begin{equation}
a_{-1}=\frac{q}{Q}\frac{X_{H}}{m^{2}}.
\end{equation}
Thus both components diverge near the horizon, in accordance with (\ref{ao-r}%
,\ref{ao-t}), while satisfying 
\begin{equation}
a^{(r)}_o+a^{(t)}_o=O(N).  \label{rt-sum}
\end{equation}

\subsection{Falling observers}

The falling frame $e^{(i)}_f$ is attached to a particle falling into the
black hole according to (\ref{rrt})--(\ref{va}), with the Lorentz factor $%
\gamma_{f}=1/N$ and velocity $v_f =1-O(N^2)$ in the static frame.

In this frame the tetrad components of acceleration are equal to 
\begin{align}
m a^{(t)}_f&=-\gamma_F \;\frac{qQ}{r^{2}}\,\frac{Z-v_F X}{mN}; \\
m a^{(r)}_f&=+\gamma_F \;\frac{qQ}{r^{2}}\,\frac{X-v_F Z}{mN}.
\end{align}

For usual particles, in the horizon limit $N\rightarrow 0$, $X_{H}\neq 0$.
Then (\ref{va}) and (\ref{zz}) imply that $Z-X=O(N^{2})$, so $a^{(t)}_f$ and $a^{(r)}_f$
are finite.

If the particle under consideration is critical, then $Z\sim X\sim N$, and
both components of acceleration diverge: 
\begin{align}
a^{(t)}_f=&\frac{\tilde{a}_{-1}}{N}+O(N), \\
a^{(r)}_f=&\frac{\tilde{a}_{-1}}{N}+O(N), \\
&\tilde{a}_{-1} =r_H^{-1}\,\frac{q}{m} \Big[\frac{q}{m}-\sqrt{%
\frac{q^2}{m^2}-1}\Big].
\end{align}
Thus we see that, indeed, all the general properties (\ref{af-t}), (\ref{af-r}), (\ref{ao-t}), (\ref{ao-r}), described in the preceding section, are
explicitly verified in this exactly solvable case.

\section{BSW effect under finite forces: equatorial motion}

\subsection{Motion in equatorial plane}

Let $m=1$, and let us consider motion in the equatorial plane so that $%
u^{z}=0$ and $a_{o}^{(z)}=0$. Then for arbitrary motion we have 1) the
normalization condition for velocity 
\begin{equation}
u^{r}=-\frac{\sqrt{A}}{N}Z,\qquad Z=\sqrt{X^{2}-N^{2}\Big[\frac{L^{2}}{%
g_{\phi }}+1\Big]},  \label{zeq}
\end{equation}%
and 2) orthogonality condition for acceleration, which can be written in terms of (\ref{a}--\ref{a-z}) as 
\begin{align}
0=u_{\mu }a^{\mu }& =+u_{t}a^{t}+u_{\phi }a^{\phi }+u_{r}a^{r} \\
& =-Ea^{t}+La^{\phi }+A^{-1}u^{r}a^{r} \\
& =-\frac{X}{N}a_{o}^{(t)}+\frac{L}{\sqrt{g_{\phi }}}a_{o}^{(\phi )}
	+\frac{u^{r}}{\sqrt{A}}a_{o}^{(r)}.  \label{Orth3a}
\end{align}%
Generically, at least two of the three components of $a^{(i)}$ have to be
non-zero if there is acceleration. Also for simplicity we will assume\footnote{This assumption is purely technical. In general, one should write $A=N^2 B$, where $B$ is some bounded function which does not vanish at the horizon. Its form does not affect the results qualitatively but leads to 
more cumbersome expressions. Thus we put for simplicity $B=1$, which also
fixes the time scale.} that $A=N^{2}$, so that $u^{r}=-Z$ and orthogonality condition takes form
\begin{equation}
\frac{X}{N}a_{o}^{(t)}-\frac{L}{\sqrt{g_{\phi }}}a_{o}^{(\phi )}
	+\frac{Z}{N}a_{o}^{(r)}=0.\label{Orth3}
\end{equation}

Of the four components of the equation of motion 
\begin{equation}
(u^\mu \nabla_\mu) u^\nu=a^\nu
\end{equation}
one is trivial\footnote{In the equatorial plane derivatives of metric functions by $z$ in (\ref{aZ}) must vanish due to symmetry.} $a^{z}=0$, and the other three are related through the
orthogonality condition, so it is always sufficient to consider only two
components, for example (\ref{dEdt}) and (\ref{dLdt}), which can be written
in terms of $X$ and $L$ as 
\begin{align}
\frac{dX}{d\tau}&=N a_o^{(t)}-L\frac{d\omega}{d\tau};  \label{Xeq} \\
\frac{dL}{d\tau}&=\sqrt{g_\phi}\; a_o^{(\phi)} .  \label{Leq}
\end{align}
As $dr/d\tau =u^r =-Z$, in terms of $X$ and derivatives by $\xi\equiv (r-r_H
)$, which are denoted by primes, this can be written as 
\begin{align}
&X^{\prime}+L\omega^{\prime}=-\frac{N}{Z} a_o^{(t)};  \label{Xeq2} \\
&L^{\prime}=-\frac{\sqrt{g_\phi}}{Z}\; a_o^{(\phi)} .  \label{Leq2}
\end{align}
It can be checked that, indeed, in case $u^z =0$, equations (\ref{Xeq}) and (%
\ref{Leq}) together with (\ref{Orth3}) give (\ref{aR}).

\subsection{Acceleration in proper frame}
Expressing acceleration components in the OZAMO frame through the particle's parameters $E$ and $L$ from (\ref{Xeq2},\ref{Leq2}) and the orthogonality condition (\ref{Orth3}), one obtains
\begin{align}
a_o^{(\phi)}&=-\frac{Z}{\sqrt{g_\phi }}L' ;\label{EQao-phi}\\
a_o^{(t)}&=-\frac{Z}{N}(X'+L\omega');\label{EQao-t}\\
a_o^{(r)}&=-\frac{X}{Z}a_o^{(t)}-N\frac{LL'}{g_\phi}.\label{EQao-r}
\end{align}
For critical particles the OZAMO frame is the proper frame. For other types of particles the $r$ and $t$ components of acceleration in the proper frame are given by (\ref{AProper}) with Lorentz factor (\ref{gx}) 
\begin{equation}
\gamma =\frac{X}{N},  \label{gx2}
\end{equation}
while $a_{pr}^{(\phi)}=a_{o}^{(\phi)}$ for any type. Using (\ref{EQao-phi}--\ref{EQao-r}), this gives
\begin{equation}
\begin{pmatrix}
a_{pr}^{(t)}\\a_{pr}^{(r)}
\end{pmatrix}=
\frac{X}{N}\left\{
	\frac{a_o^{(t)}}{Z}
	\begin{pmatrix}
		Z-vX \\ Zv-X
	\end{pmatrix}
	-N\begin{pmatrix}
		v\\1
	\end{pmatrix}
		\frac{LL'}{g_\phi}\right\}. \label{ACCproper}
\end{equation}

Suppose we have a particle with
\begin{align}
	X&=\alpha \xi^p (1+O(\xi)),\qquad 	\gamma=\frac{X}{\xi},\\
	v&=\sqrt{1-\gamma^{-2}}=1-\frac{1}{2\alpha}\xi^{2(1-p)}(1+O(\xi)),
\end{align}
where $p<1$. Using (\ref{Z-X}), we get
\begin{align}
Z-Xv&=(\tfrac12 -C)\xi^{2-p}(1+O(\xi)), \label{Z-Xv}\\
X-Zv&=(\tfrac12 +C)\xi^{2-p}(1+O(\xi)).\label{X-Zv}
\end{align}

\subsubsection{Usual particles}
For a usual particle $p=0$, so assuming $L'$ is bounded, $a_o^{(\phi,r,t)}=O(1)$, while $\gamma \sim 1/\xi$, and one can easily check term by term that 
acceleration in the proper frame (\ref{ACCproper}) is always bounded: as expected, \textbf{for usual particles there are no additional requirements}.

\subsubsection{Sub-critical particles}
For a sub-critical particle $p\in(0,1)$. As $X\sim Z$, the derivative $X^{\prime}\sim \xi^{p-1}$ in (\ref{EQao-t}) diverges, while $L\omega^{\prime}=O(1)$, so 
\begin{equation}
\frac{a_o^{(t)}}{Z}\sim \xi^{p-2}.
\end{equation}
Then taking into account (\ref{Z-Xv}--\ref{X-Zv}), the first term in the braces of (\ref{ACCproper}) is $O(1)$, and different in the two rows, thus separated from zero. 

The second term could only compensate the first one (in one of the two rows), if $L' \sim \xi^{-1}$, which would imply divergent $L\sim \ln \xi$. Therefore the quantity in the braces is finite and separated from zero, so the proper acceleration diverges as (\ref{gx2})
\begin{equation}
\gamma =\frac{X}{N}\sim \xi^{p-1}\to \infty .
\end{equation}
This means that there are \textbf{no sub-critical particles} with finite acceleration for motion in the equatorial plane.

\subsubsection{Critical particles}
The only remaining case to be considered is critical particles. Although (\ref{ACCproper}) for them is unnecessary, one restores the acceleration in the OZAMO frame from it by setting $\gamma=1$ and $v=0$. We see that\footnote{Remember that components $a_o^{(i)}$ are related through the orthogonality condition (\ref{Orth3}); if two of them are finite, then the third is bounded as well.}
\begin{align}
	a_o^{(\phi)}&\sim \xi L' ;\\
	a_o^{(r,t)}&\sim (X'+L\omega '),
\end{align}
so in order for such trajectory to be realized we need the azimuthal force to tend to zero fast enough:
\begin{equation}
a_o^{(\phi)}=O(\xi).
\end{equation}
There is no restriction on the radial component: it can be of the order of unity, as it will still be possible to fine-tune a critical particle by the appropriate choice of initial condition (this will be shown in more detail in the next Section). Thus the radial component does not affect or hinder the existence of critical trajectories and consequently the BSW effect.  This is in agreement with the already established fact that the radial force itself is the reason for the BSW effect near charged nonrotating black holes \cite{jl}.

\subsection{Example: azimuthal dissipative force}
Let us consider the particular case when the radial force, which does not hinder critical particles anyway, is absent:
\begin{equation}
a_{o}^{(r)}= 0,\qquad a_{o}^{(t)}, a_{o}^{(\phi)}\neq 0 .
\end{equation}
Using orthogonality (\ref{Orth3}), 
\begin{equation}
a_o^{(t)}=\frac{N}{X}\frac{L}{\sqrt{g_\phi}} a_o^{(\phi)} ,  \label{OrthAD}
\end{equation}
so in terms of derivatives with respect to $\xi$ Eqs. (\ref{Xeq2}) and (\ref{Leq2}) can be rewritten as 
\begin{align}
&g_\phi X(X^{\prime}+L\omega ^{\prime})=N^2 LL^{\prime};  \label{AD-eq1} \\
&a_o^{(t)}=-N \frac{Z}{X}\frac{LL^{\prime}}{g_\phi} ;  \label{AD-eq2}\\
&a_o^{(\phi)}=-Z\frac{L^{\prime}}{\sqrt{g_\phi}}.\label{AD-eq3}
\end{align}

\subsubsection{Tuning a critical particle}
In this section we show in more detail how one would tune the particle to be
critical $X\sim \xi$.

Assuming expansions 
\begin{align}
N^2 &=\nu_2 \xi^2 +\nu_3 \xi^3 +\ldots , \\
\omega&=\omega_H -\omega_1 \xi +\omega_2 \xi^2 +\ldots , \\
g_\phi &=g_H +g_1 \xi +g_2 \xi^2 +\ldots , \\
X&=x_1 \xi +x_2 \xi^2 +\ldots , \\
L&=l_H + l_1 \xi +l_2 \xi^2 +\ldots ,
\end{align}
from (\ref{AD-eq1}) we obtain in consecutive orders 
\begin{align}
l_H &=\frac{x_1}{\omega_1}; \\
l_{1}&=2\frac{x_1 \omega_2 +x_2 \omega_1}{\omega_1^2 +\nu_2 /g_H}; \\
l_{2}&=l_{2} (x_1, x_2, x_3),\quad \ldots
\end{align}

Then 
\begin{equation}
\frac{Z^{2}}{\xi^2}
	\approx x_1^2 \Big[1-\frac{\nu_2}{g_H}\omega_1^{-2}\Big]-\nu_2 .
\end{equation}
There is a critical particle for 
\begin{equation}
|x_1| >x_{1\min},
\end{equation}
and there is a solution 
\begin{equation}
x_{1\min}=\frac{\nu_2}{1-\frac{\nu_2}{g_H}\omega_1^{-2}}
\end{equation}
as long as 
\begin{equation}
\omega_1^2 > \frac{\nu_2}{g_H}.  \label{OmegaCond}
\end{equation}

From (\ref{AD-eq2}) we get 
\begin{equation}
a_o^{(t)}=-\frac{Z}{N}(X^{\prime}+L\omega ^{\prime})
\end{equation}
and after substitution of expansions for $X,L$ and $\omega$, 
\begin{equation}
a_o^{(t)}\approx -\frac{2Z}{\omega_1 \sqrt{\nu_2}}\frac{x_1 \omega_2 +x_2
\omega_1}{1+g_H \omega_1^2 /\nu_2}\sim Z \sim \xi .
\end{equation}
Further terms are obtained straightforwardly but they are quite cumbersome.

So, if acceleration is expanded in a series by $\xi$ 
\begin{equation}
a_o^{(t)}=a_1 \xi +a_2 \xi^2 +\ldots,
\end{equation}
in the first order we obtain $a_{1}(x_1 ,x_2)$. As long as $a =O(\xi)$, and the
metric coefficients satisfy (\ref{OmegaCond}), we can take arbitrary $x_{1}$
such that $|x_1|>x_{1\min}$ (or equivalently $l_{H}=x_{1}/\omega_{1}$). Then
for the given $a_1$ in the first order we obtain $x_{2}(a_1)$, in the next
order $x_3 (a_1, a_2)$ and so on. The set of critical trajectories\footnote{For large enough $|x_1|$ the turning point, given by $Z=0$, will be at finite values of $(r-r_H )$ from the horizon. Thus it will be at the coordinate distance that does not have to be small in order to gain arbitrarily large $E_{c.m.}$ at the collision event near the horizon. This is in contrast to the case discussed in \cite{BSW2}, which is realized near the turning point of a usual particle with small $X_H$, and the turning point itself must be close to the horizon: the nearer it is, the larger $E_{c.m.}$ is achieved.} is parametrized by one free parameter $x_1$ (or $l_H$).

\subsubsection{Other realizations of critical trajectories}
Suppose now the azimuthal force tends to zero as $\xi^s$ with some integer $s>1$. Then from (\ref{AD-eq3}) we see that $L' \sim \xi^{s-1}$, and therefore expansion (\ref{el}) for $L$ near the horizon takes the form
\begin{equation}
L=L_H +L_{s}\xi^{s}(1+o(1)).\label{LS}
\end{equation}
It is perfectly consistent with the particle being critical, so that $X\sim \xi$: (\ref{AD-eq1}) can be satisfied for any integer $s>1$ and solved for $X(L)$ (or, equivalently, $E(L)$) in each consecutive order by $\xi$. Let us consider, for example, the case $s=2$. Assuming
\begin{equation}
X=x_1 \xi +x_2 \xi^2 +x_3 \xi^3 +O(\xi^4),
\end{equation}
from (\ref{AD-eq1}) in consecutive orders one obtains
\begin{align}
x_{1}&=L_{H}\omega_{1};\\
x_{2}&=-2\omega_{2}L_{H};\\
x_{3}&=-\omega_3 L_H +\frac{\omega_1}{3}L_2 +\frac{2}{3g_{\phi H}}L_H L_2 ;\\
\ldots \notag
\end{align}
This can be turned around to give $L_H (x_1)$ and $L_2 (x_3)$, but $x_2 / x_1$ is fixed to metric function coefficients; in terms of $E(L)$ this is
\begin{equation}
E=\omega_H L_H +(L_2 \omega_H -L_H \omega_2)\xi^2 +\ldots .
\end{equation}
For other integer $s$ the procedure is analogous.

\section{Energy bounds in collisions with near-critical particles}
We have seen in the previous section that as long as the azimuthal force is weak enough, critical particles exist and can be tuned via initial conditions. Then the BSW effect in its primary version \cite{ban} manifests itself. However, what if this is not the case and azimuthal force is e.g. separated from zero on the horizon? The condition for critical particles is that
\begin{equation}
a^{(\phi)} =O(x) \label{Need},
\end{equation}
where
\begin{equation}
x=\frac{\xi}{r_H}  \label{Xdimensionless}
\end{equation}
is the dimensionless radial coordinate. Suppose that instead
\begin{equation}
a^{(\phi)}(x)\approx \frac{a_{0}x ^{\lambda }}{r_H},\qquad \lambda <1 ,  \label{Asmall}
\end{equation}
so that (\ref{Need}) is violated, and we factored out the dimensional quantity $r_H^{-1}$, so that $a_0$ is dimensionless.

Does it mean that the BSW effect necessarily breaks down? Under no
additional assumptions -- yes. However, what if $a_0$ is small? Radiation
reaction forces are usually considered very small (see e.g. \cite{ne}, \cite{Kesden}). In that case, the question is how high $E_{c.m.}$ can be
achieved for the given small $a_{0}$ ? 

Let us reformulate the condition that \emph{is} satisfied (\ref{Asmall}) via
another small parameter: 
\begin{equation}
a^{(\phi)} (x)\approx r_H^{-1} \; x^{\lambda}x_{m}^{1-\lambda},
\end{equation}
where 
\begin{equation}
x_{m}=a_0^{\frac{1}{1-\lambda}}\ll 1 .
\end{equation}
Then 
\begin{equation}
a^{(\phi)}(x_m) \approx r_H^{-1} x_m
\end{equation}
and for all $x\gtrsim x_m$ the necessary condition for acceleration (\ref{Need}) is
effectively obeyed.

But then for $\xi\gtrsim r_H x_m$ the trajectory of a particle can be
effectively tuned to be critical, (or sub-critical, for the chosen $p$), at
will. At the near-horizon end of this region, i.e. at $\xi_{m}\sim r_H x_m$,
the Lorentz factor with a usual particle with $p_2 =0$ will behave as
described (\ref{p1-p2}) and can grow very large. Thus, for $p=1$ (the particle is tuned to be critical) from (\ref{gamma12-1}) we get 
\begin{equation}
\gamma_{12}^{(max)} \approx 
	\gamma_{12}(x_m) \approx 
		\frac{C_1 \alpha_2}{r_H} \cdot a_0^{-\frac{1}{1-\lambda}}.
\end{equation}

We see that as long as the amplitude $a_0$ of the azimuthal force acting on the particle is small enough, the BSW effect survives almost any
kind of perturbation: one only has to calculate accurately the corresponding tuning parameters for the effectively critical trajectories.

\section{Kinematic restrictions on critical particles and two types of the
BSW effect}

In the preceding Section, it was assumed that collision occurred not exactly
on the horizon but at some coordinate distance from it, its scale being tied
to the amplitude of azimuthal acceleration, which is supposed to be small. The force, being too large, prevents
the critical particle from approaching the horizon. In this sense, the
reason of it is dynamic. Meanwhile, even if the external force is small enough or absent at all, pure
kinematic factors can also create an obstacle for reaching the horizon.

Let us remind the situation with geodesic particles. If the horizon is
nonextremal, the critical particle cannot reach the horizon at all.
Nonetheless, it was demonstrated in \cite{gp} for the Kerr metric and in 
\cite{prd} for generic dirty axially symmetric black holes, that $E_{c.m.}$
can be made as large as one likes provided (i) the critical particle is
replaced with a slightly noncritical one, (ii) the coordinate distance
between the point of collision and horizon is adjusted to the small
deviation of the particle's parameters from the values corresponding to the
critical case.

Now, we are dealing with an extremal horizon but there is a special situation 
when $X\sim \xi^p$ with $p>1$ (``supercritical'' particle). Such a particle cannot reach the horizon (in this sense it is similar to the case of the nonextremal horizon).

Then, instead of taking a critical particle, we can choose a usual one with sufficiently
small $X_{H}$. More precisely, let us consider expansion for $X$ of the form 
\begin{equation}
X=X_{H}+\alpha_{s}\xi ^{s}(1+x_{1}\xi +\ldots ),\qquad p>1.  \label{xpp}
\end{equation}

Then, we look for the region in which both terms in $Z$ (\ref{z}) are of the
same order of magnitude. This is achieved at $\xi\approx \xi_c \sim r_H X_H$. Then we can neglect the correction in (\ref{xpp}), so that 
\begin{equation}
Z(\xi_c)\sim X_{H}\sim \xi_c \sim N(\xi_c ),
\end{equation}%
and therefore (\ref{gacm}) implies 
\begin{equation}
\gamma _{c.m.}^{(max)}
	\approx \gamma_{c.m.}(\xi_c)
		\sim N^{-1}(\xi_c) 
		\sim \xi_{c} ^{-1} .  \label{bsw}
\end{equation}

Thus one can distinguish between two main types of the BSW effect: BSW 1, in
which the critical particle can approach the horizon, so that the
horizon limit can be taken, and BSW 2, for which the critical
particle does not reach the horizon. We see that, in general, the presence
of the external force is compatible with both types of the BSW effect\footnote{Another, more detailed classification of trajectories and corresponding
types of the BSW effect can be found in Sec. IV of \cite{ne} for the Kerr
metric and in \cite{jh} for general dirty rotating axially symmetric black
holes.}.

It is worth noting that even in the absence of force, the expansion for $X$
can take the form (\ref{xpp}), if the linear terms cancel each other. Say,
this happens for solutions near the so-called ultraextremal horizon ($%
N^{2}\sim (r-r_{H})^{3}$) in special "exotic" metrics described in
Sec.\,IV\,B.\,5 of \cite{reg}, for which $\partial_r \omega \big|_H
=\partial^2_r \omega \big|_H =0$, and thus $s$ can be equal to $2$ or $3$.
Correspondingly, the BSW-2 effect can be realized near such horizons.

\section{Conclusion}

In general, three main circumstances were considered as the factors which
were expected to restrict the indefinite growth of $E_{c.m.}$ and thus
create obstacles to the manifestation of the BSW effect. These are (i)
self-gravitation, (ii) deviation of a black hole from extremality \cite{ted}%
, and (iii) the force due to backreaction of gravitational or
electromagnetic radiation. As far as self-gravitation is concerned, it was
shown in \cite{shell} that for collisions of massive shells, either the BSW
effect does not occur or it occurs but in the region inaccessible by a
remote observer. However, in this case the shell does not approach the
horizon from the viewpoint of an external observer. As there is no horizon,
there is no BSW effect. Factor (ii) was analyzed in \cite{gp} for the Kerr
metric where it was shown that for nonextremal black holes the BSW effect
does exist (this conclusion was generalized in \cite{prd} to generic dirty
axially symmetric black holes). 

And, in the present work, we showed for extremal horizons that the BSW effect is compatible with a nonzero force under rather general assumptions: the radial force should be finite and the azimuthal force should tend to zero not too slowly. In terms of energy and angular momentum, the kinematic condition for the realization of the BSW effect is the same as for geodesic particles: $E=\omega_H L$. In this sense, this condition by itself survives the action of the force (see also Sec. V of \cite{insp} where, however, another physical situation was considered -- near-circular orbits around near-extremal black holes). Our approach is model-independent and is based on general properties of the horizons. 

For the finite radial component of the force and the azimuthal one which tends to zero near the horizon as $r-r_H$, the BSW effect still exists. Otherwise, the effect is formally absent. The reason for the restriction on the azimuthal force seems to be clear: if azimuthal force was too large or did not tend to zero, in the infinite proper time that it takes for a critical particle to reach the horizon this force would accelerate the particle to infinite values of angular momentum. Obviously, one would not expect a force dissipative by nature, such as radiation reaction, to have such an effect. So, this only seeming restriction should be always obeyed. Even if it is not (so critical trajectories are absent), but the amplitude of the azimuthal force is small, the restrictions on $E_{c.m.}$ are shown to be inessential, and one can still attain very high energies. 

To summarize, the BSW effect turned out to be more viable than one could expect.

The present work confirmed that the BSW effect relies on two main
properties: (i) the presence of the horizon and (ii) the existence of
special types of trajectories. Thus it has geometric nature and reflects
general features of black holes irrespective of the details of the system.
Concrete realization of the BSW effect certainly depends on particular
properties of a system but near the horizon these properties manifest
themselves in a universal way. We see that although dissipative forces in flat spacetime generically bound the values of energy peaks from above, in the strong gravitational field regime near the horizon the geometry dominates over the influence of dissipative forces on the system.

The present results refer to extremal horizons only. The nonextremal case
and, especially, motion on circular orbits around near-extremal black holes,
so important in astrophysical context, require separate treatment. Generalization to non-equatorial motion is also necessary. This will
be done elsewhere.

\end{document}